\documentclass[prl,superscriptaddress, twocolumn, preprintnumbers,amssymb]{revtex4}
\usepackage[pdftex]{graphicx}
\usepackage{amsmath,amsthm,amssymb,mathtools}
\usepackage[colorlinks,linkcolor=blue]{hyperref}
\usepackage[all]{hypcap}

\usepackage{xcolor}

\newcommand*{\bR}{{\bf R}}
\newcommand*{\br}{{\bf r}}

\begin{document}
\title{
Competing collinear magnetic structures in superconducting FeSe by first principles quantum Monte Carlo calculations}
\author{Brian Busemeyer}
\affiliation{Department of Physics, University of Illinois at Urbana-Champaign}
\author{Mario Dagrada}
\affiliation{IMPMC, Sorbonne Universit\'es, Universit\'e Pierre et Marie Curie, CNRS, IRD, MNHN, 4 place Jussieu, 75252 Paris, France}
\author{Sandro Sorella}
\affiliation{SISSA}
\author{Michele Casula}
\affiliation{IMPMC, Sorbonne Universit\'es, Universit\'e Pierre et Marie Curie, CNRS, IRD, MNHN, 4 place Jussieu, 75252 Paris, France}
\author{Lucas K. Wagner}
\affiliation{Department of Physics, University of Illinois at Urbana-Champaign}
\date{\today}

\begin{abstract}
Resolving the interplay between magnetic interactions and structural properties in strongly correlated materials through a quantitatively accurate approach has been a major challenge in condensed matter physics. 
Here we apply highly accurate first principles quantum Monte Carlo (QMC) techniques to obtain structural and magnetic properties of the iron selenide (FeSe) superconductor under pressure. Where comparable, the computed properties are very close to the experimental values.
Of potential ordered magnetic configurations, collinear spin configurations are the most energetically favorable over the explored pressure range. 
They become nearly degenerate in energy with bicollinear spin orderings at around 7 GPa, when the experimental critical temperature $T_c$ is the highest. 
On the other hand, ferromagnetic, checkerboard, and staggered dimer configurations become relatively higher in energy as the pressure increases. 
The behavior under pressure is explained by an accurate analysis of the charge compressibility and the orbital occupation as described by the QMC many-body wave function, which reveals how spin, charge and orbital degrees of freedom are strongly coupled in this compound. 
This remarkable pressure evolution suggests that stripe-like magnetic fluctuations may be responsible for the enhanced $T_c$ in FeSe and that higher T$_c$ is associated with nearness to a crossover between collinear and bicollinear ordering.
\end{abstract}

\maketitle

\section{Introduction}

The quest for a microscopic theory of unconventional or high-temperature superconductivity is a major challenge in condensed matter physics. The discovery of iron-based superconductors in 2006~\cite{kamihara_iron-based_2006} was an important contribution to the field since it added a second class of high-temperature unconventional superconductors to the experimental roster, 
along with the cuprate superconductors.
Despite their different electronic structure, their phase diagrams have striking similarities\cite{norman_high-temperature_2008,demedici_ibs},
particularly the proximity of the superconducting phase with an antiferromagnetic state. 
This behavior, along with other considerations\cite{scalapino_common_2012,wen_interplay_2011,dai_antiferromagnetic_2015,dai_magnetism_2012,imai_why_2009}, 
makes it likely that spins and magnetism are important in determining the superconducting state. 
 
FeSe is a particularly interesting example of the iron-based superconductors for several reasons.
Its critical temperature is strongly dependent on pressure\cite{mizuguchi_superconductivity_2008,imai_why_2009,margadonna_pressure_2009}, reaching 37 K at 7 GPa.
At ambient conditions, FeSe has a simple P4/nmm crystal structure 
with two inequivalent Fe and Se positions per unit cell, 
and it undergoes a distortion from tetragonal to orthorhombic symmetry by cooling it down below 91 K, while it becomes superconductors below 8 K at ambient pressure.
An intriguing peculiarity of FeSe is that, at variance with most of the iron-based superconductors, it does not
show any long-range magnetic order over the whole phase diagram\cite{ordering}. In spite of this, very strong antiferromagnetic
spin fluctuations have been revealed by neutron scattering experiments (see for example Ref.~\onlinecite{neutron}) 
in the proximity of the superconducting phase. 
Their role in driving the nematic transition and their connection to superconductivity have been the subject of intense debate.
All these aspects make it attractive for computational techniques to correlate microscopic electronic structure 
with the superconductivity and it is therefore one of the most studied iron-based superconductors. 
However, the precise calculation of the properties of this material remains challenging from first principles 
methods such as density functional theory (DFT) due to strong electron correlation.

For example, the PBE band structure is in poor agreement with experiments which report a considerably narrower bandwidth\cite{Borisenko2014,Audouard2015}.
Furthermore the FeSe lattice constants display an average error of $\sim$ 0.1 \AA~ independently from the exchange correlation functional employed (see for instance Ref~\cite{FeCh_transport} and Tab~\ref{tab:geometry}). 
Despite useful work using dynamical mean field theory\cite{Liebsch2010,Yin2011,aichhorn_theoretical_2010,Craco2010,Craco2014,Leonov2015} and GW\cite{gw_kotliar_2012,Imada2013,Imada2015} methods, 
there is a strong need for high quality calculations that can better describe the electronic and crystal structure of these materials.

In this article, we describe the results of first principles quantum Monte Carlo simulations of the magnetic behavior of FeSe under pressure.
The main method used in this article, fixed node diffusion Monte Carlo (FN-DMC), has been shown recently to offer very accurate results on a number of challenging materials, including VO$_2$\cite{zheng_computation_2015}, cuprates~\cite{wagner_ground_2015,kent_cuprates2014}
and other transition metal oxides as well as rare earths as cerium~\cite{CeriumQMC}. Furthermore, a recent work~\cite{micheleFeSe}, 
based on quantum Monte Carlo techniques, successfully tackled the problem of pairing symmetry in FeSe itself.

We find that, compared to commonly used density functional theory calculations, the FN-DMC calculations obtain more accurate lattice constants, bulk moduli, and band dispersion.
By increasing the pressure, the difference in energy of ordered magnetic states with stripe-like order goes to zero with pressure, while checkerboard-like magnetic states increase in energy.
The convergence of the stripe-like magnetic states is correlated with the increase in T$_c$ in this material under pressure, which offers a tantalizing connection to spin fluctuations as a potential origin.
Such behavior may be a calculable design principle for new unconventional superconducting materials.

\section{Methods}

\subsection{Fixed-node diffusion Monte Carlo}
\begin{figure*}
  \begin{tabular}{cccc}
  \includegraphics[width=0.5\columnwidth]{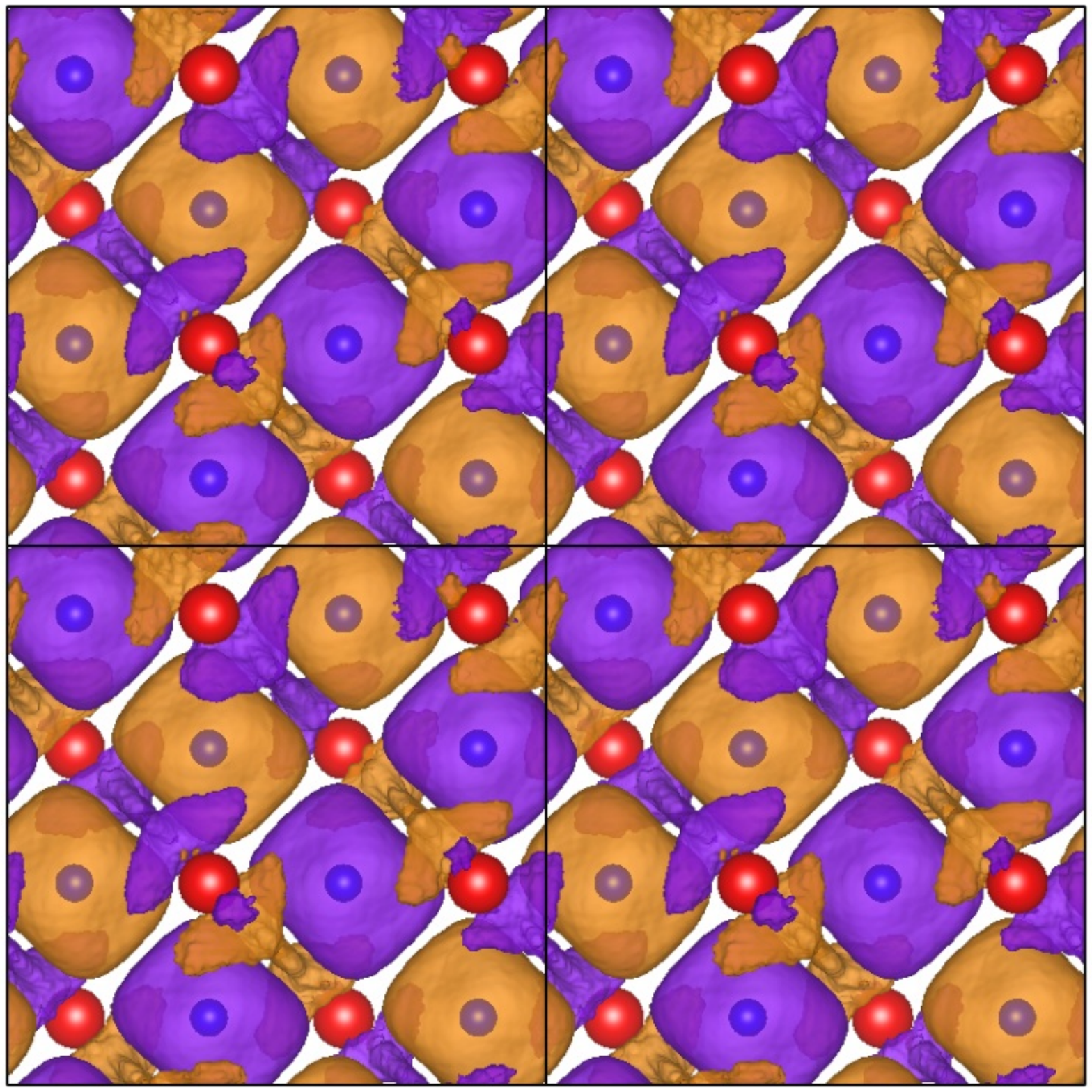} &
   \includegraphics[width=0.5\columnwidth]{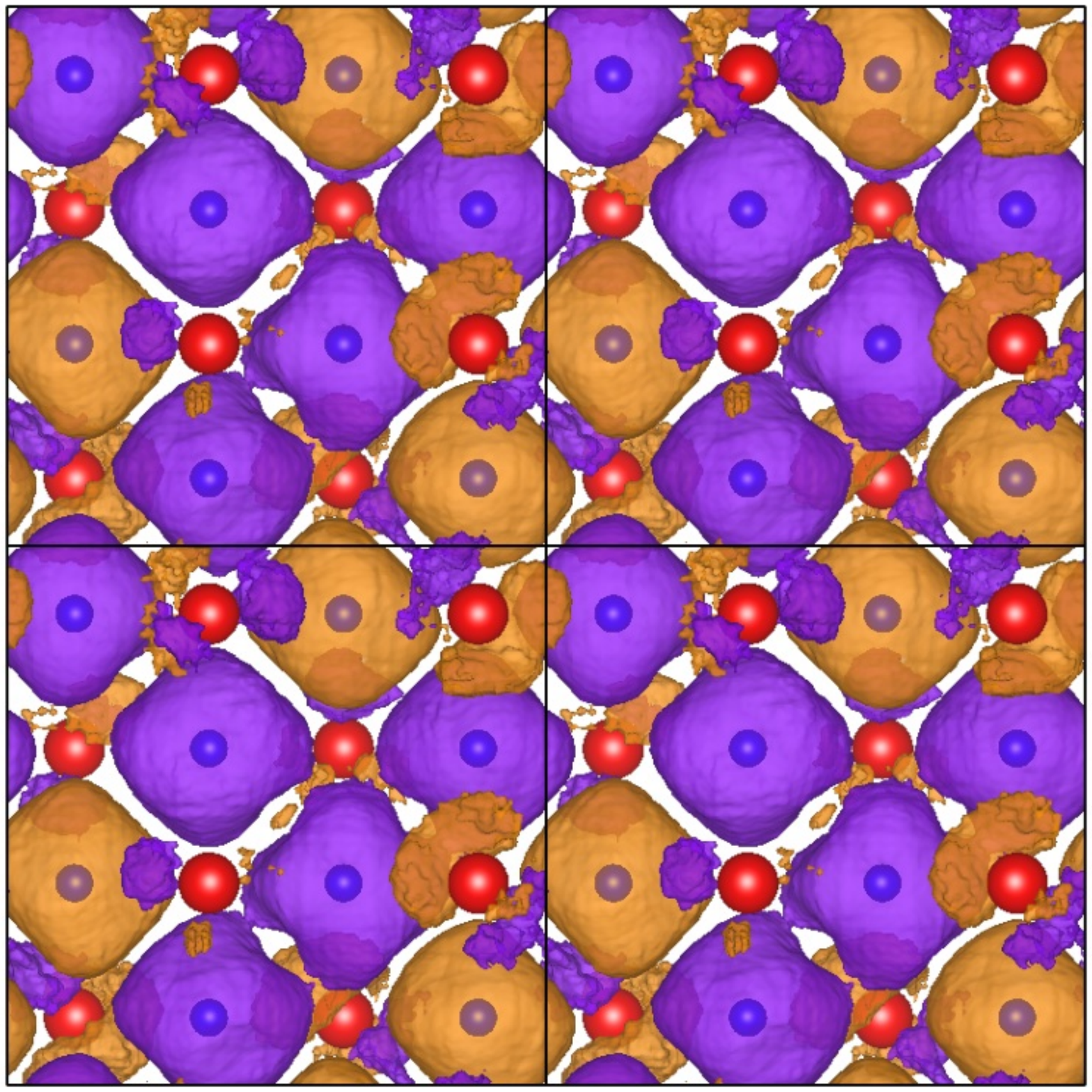} &
  \includegraphics[width=0.5\columnwidth]{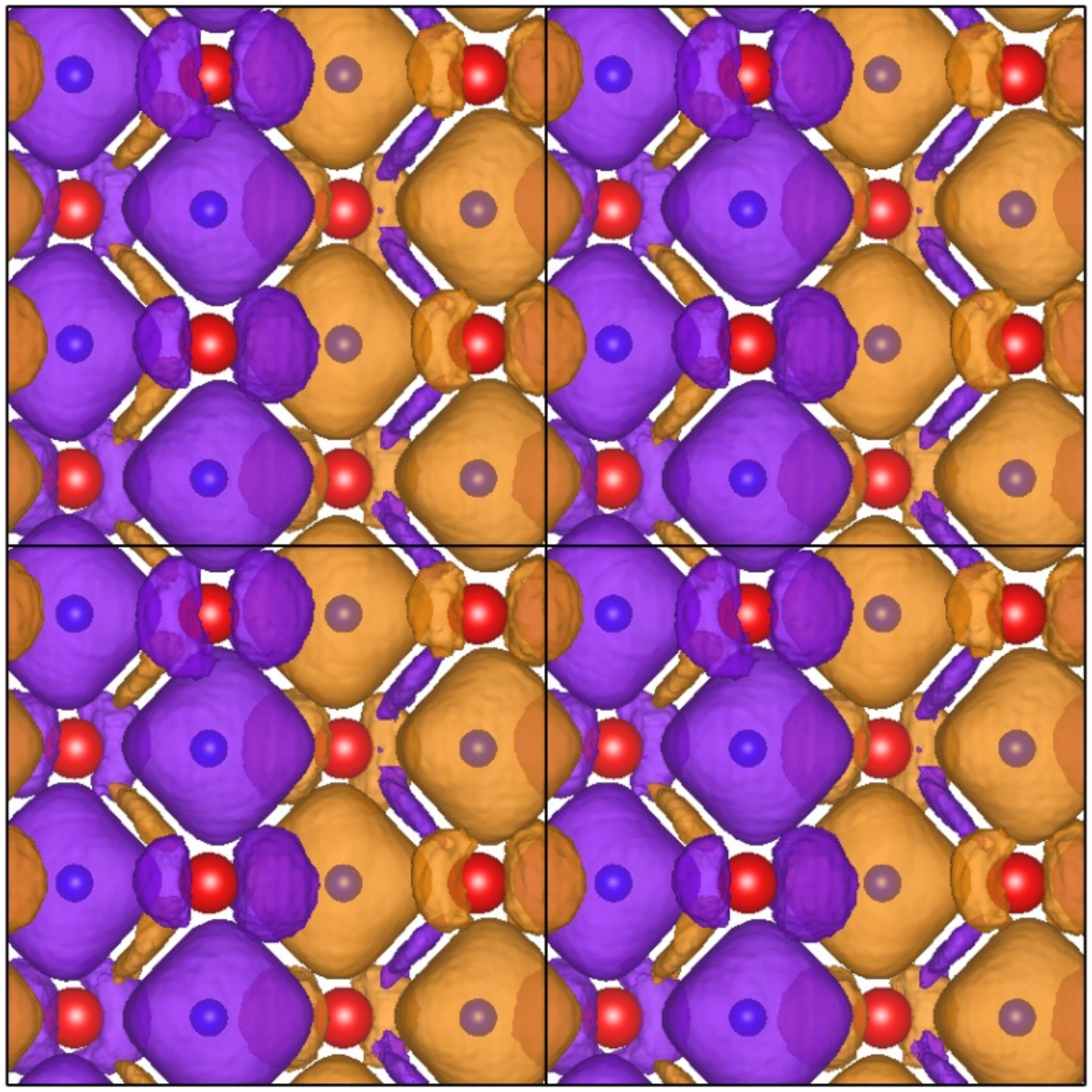} & 
   \includegraphics[width=0.5\columnwidth]{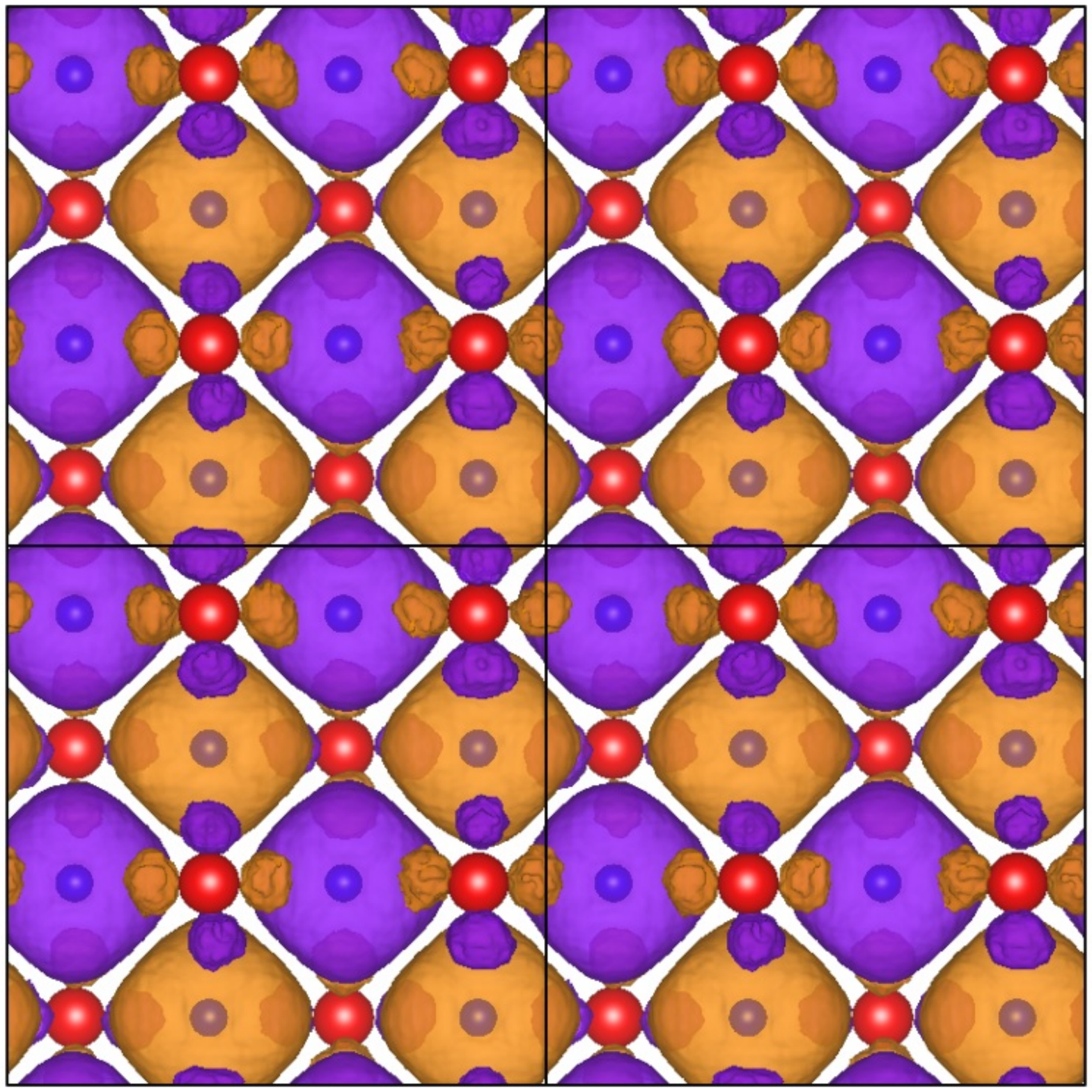} \\
   (a) & (b) &
   (c) & (d) \\
  \end{tabular}
  \caption{
    \label{fig:spinden} Spin densities of magnetic orderings at ambient
    pressure, (a) collinear, (b) collinear, 1 flip, (c) bicollinear, and (d)
    checkerboard. 
    Four unit cells of a single iron layer are shown, divided by
    black lines. ``Collinear, 1 flip" refers to flipping the spin of one
    iron per unit cell in the collinear configuration. Since four unit cells are
    shown above, there are four ``flipped" iron moments shown in this plot. The
    larger red Se atoms lie above and below the plane 
    and 
    show significantly smaller spin density. Irons are smaller and blue, and lie
    within larger concentration of spin. The two colors of the isosurfaces
    denote density of up and down. 
  }
\end{figure*}

Fixed-node diffusion Monte Carlo~\cite{foulkes_quantum_2001} is a stochastic approach to solving the exact first-principles imaginary-time  Schr\"odinger equation. 
 FN-DMC is a variational (upper bound to the ground state) method with no adjustable parameters.
 FN-DMC is a projector-based approach, which projects out the ground-state wave function via repeated application of the 
operator
$e^{-(\hat H - E_0)\tau}$ on some trial wavefunction $|\Psi_T\rangle$:
    \begin{align*}
      |\Psi_\mathrm{DMC}\rangle
      &=
      \lim_{\tau \to \infty}
      e^{-(\hat H - E_0) \tau} |\Psi_T\rangle
      \\
      &=
      \lim_{\tau \to \infty}
      \sum_i e^{-(E_i - E_0) \tau} | \Psi_i \rangle  \langle \Psi_i | \Psi_T \rangle
      \\
      &=
      | \Psi_0\rangle  \langle \Psi_0 | \Psi_T \rangle
    \end{align*}
With no further approximation, this method is exact, however, it suffers from the well-known fermion sign problem. 
This is circumvented in FN-DMC with the fixed-node approximation~\cite{fn1,fn2}, which constrains the projected wave function to have the same nodes as the starting trial wave function. 
If the nodal structure of the trial wave function  coincides with the ground-state one, the method remains exact. 
In practice, for accurate trial wavefunctions, this approximation introduces small errors that we will estimate.

We used a single Slater-Jastrow trial wave function \textit{ansatz} as follows:
\begin{equation}
\Psi(\bR_{\textrm{el}})=\text{Det}\left[\phi_i^\uparrow(r_j^\uparrow)\right] 
       \text{Det}\left[\phi_i^\downarrow(r_j^\downarrow)\right]
       \exp\left[\sum_{\alpha,i,j} u(r_{i\alpha},r_{j\alpha},r_{ij}) \right],
       \label{eqn:sj}
\end{equation}
 where $\bR_{\textrm{el}}=\{\br_1,\br_2,\br_3,\ldots\}$,$i,j$ refer to electrons, $\alpha$ refers to nuclei, and $\uparrow/\downarrow$ indicate spin up/down respectively.
Since the FN-DMC result is determined by the nodes of the determinants in Eq.~\ref{eqn:sj}, the orbitals $\{\phi_i^{\uparrow/\downarrow}\}$ determine the degree of the fixed node approximation.  To test the effect of these orbitals, we use two approaches. 
The first optimizes a parameter in density functional theory to generate the
orbitals, and is the less computationally demanding of the two.  
The parameter to be optimized in the Slater determinant is the amount of exact
exchange $w$, for which we find the optimal value near $w
= 0.25$. This corresponds to the PBE0 functional, and this optimum is often the
case for similar calculations on transition metal systems~\cite{kolorenc_wave_2010}.
For consistency, we used $w = 0.25$ for all our calculations, and we denote this method by DMC(PBE0).
The details of the optimization of functional and of the parameter, $w$, is reported in the Supplemental Material.
The second QMC approach used in this work performs a complete variational optimization of the determinant orbitals within a relatively small basis set, and is more computationally demanding but in principle more accurate.
These calculations are denoted throughout the paper with the label QMC(opt), where QMC will refer to VMC or DMC depending on the calculation.     
Further information on the orbital optimization procedure is provided in the Supplementary Material. 
In our tests for FeSe, while the DMC(opt) technique did obtain lower energies as expected, the energy differences were consistent between the two techniques, so most data is obtained from DMC(PBE0). 
 
All DMC(PBE0) calculations were done within the open-source package QWalk~\cite{wagner_qwalk:_2009}, with orbitals generated by DFT calculations performed with the DFT code CRYSTAL~\cite{crystal}. 
For the DMC(opt) method, we used the package TurboRVB~\cite{turborvb}. 

Our only approximation to the Hamiltonian is a Hartree-Fock pseudopotential designed specifically for quantum Monte Carlo calculations~\cite{burkatzki_energy-consistent_2007,burkatzki_energy-consistent_2008}.
The energy difference between the collinear and checkerboard magnetic state is consistent between an all-electron and pseudopotential PBE0 calculation within 0.01 eV. The convergence of the most important parameters in both our QMC methods is shown in Sec.\ref{sec:convergence} of the Supplementary Material. 
For the FeSe crystal structure, the anion height above the iron planes is the only internal parameter of the compound in the tetragonal P4/nmm phase. This 
parameter represents a crucial ingredient to determine the magnetic behavior of FeSe, but its evaluation by first-principles methods is a challenging task, as detailed in Sec.\ref{crystal_structure}.
The optimization of the Se height is carried out with two different procedures. 
For the DMC(PBE0) method, the relaxed value is obtained by fitting a total energy curve with a cubic function. 
In the VMC(opt), the optimized Se height value is obtained by a direct minimization of the ionic forces within the variational Monte Carlo framework~\cite{aad}. 
By including the cell parameters in the minimization procedure, we are able to fully relax the crystal structure of FeSe at different magnetic orderings. 
We consider a minimization converged when both the forces and their error bars are lower than 10$^{-3}$ Ha/a.u. per atom.

We found that effects due to the finite size of the simulation cell, or finite size errors (FSEs), constitute the major source of systematic error for both DMC(PBE0) and DMC(opt). 
We apply several techniques in order to reduce FSEs at best. All DMC(PBE0) calculations are twist averaged~\cite{twist}
over 8 twist conditions on the 8 f.u. FeSe supercell; DMC(opt) results are instead obtained with a larger 16 f.u. supercell averaging over periodic and antiperiodic 
boundary conditions; further corrections are then applied to cure one-body and two-body FSEs. In both cases, we managed to reduce the impact of FSEs below the desired
accuracy on energy calculations. A detailed explanation of the procedures used to control FSEs is given in the Supplemental Material.

\begin{figure*}
  \includegraphics{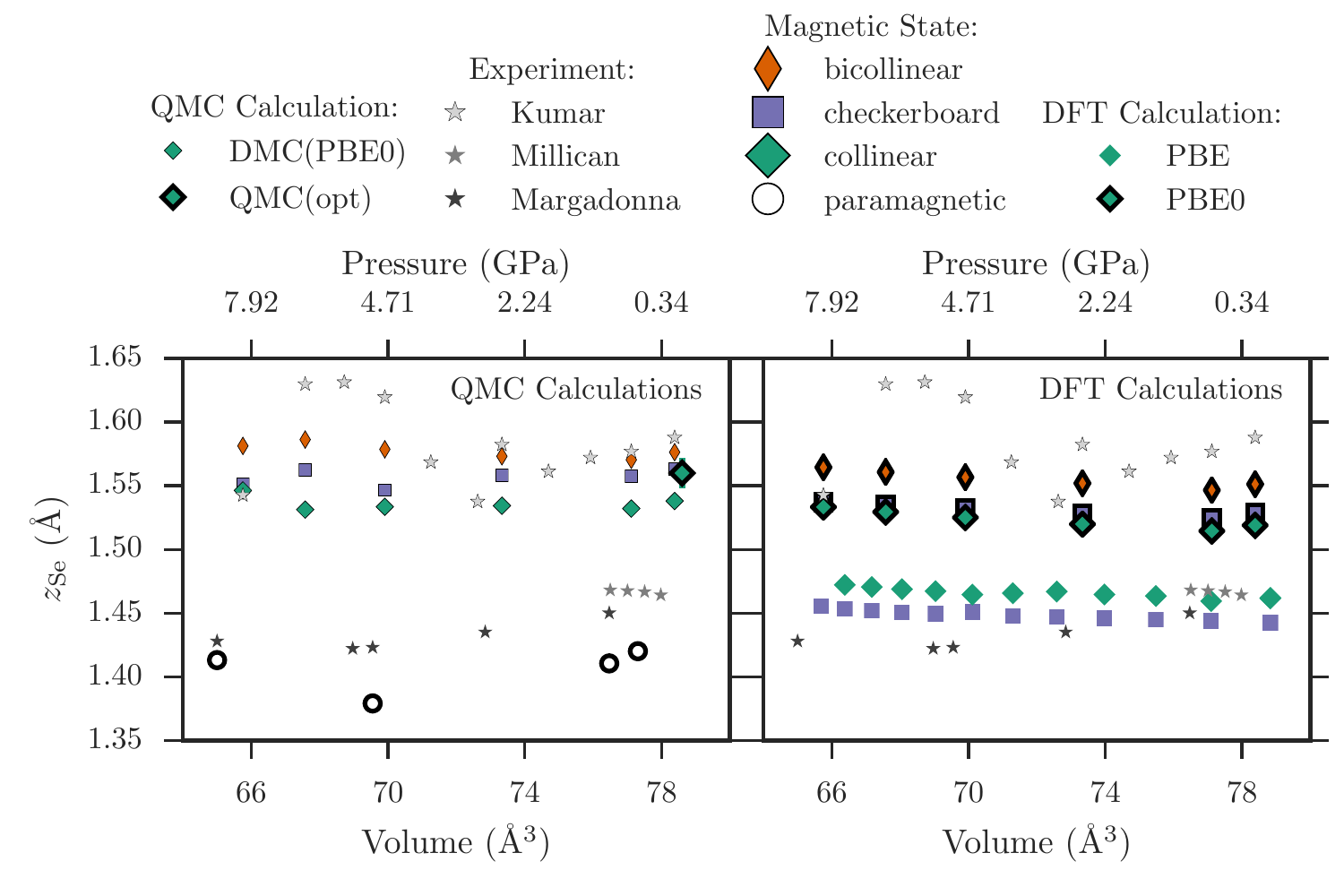}
  \caption{
    \label{fig:zse} 
    Selection of experimental measurements from Margadonna \emph{et
    al}.~\cite{margadonna_pressure_2009},
    Kumar \emph{et al}.~\cite{Kumar}, and Millican \emph{et al}.
   ~\cite{millican_pressure-induced_2009} of the selenium height,
    $z_\mathrm{Se}$, as a function of pressure, along with corresponding QMC
    (left) and DFT (right) predictions. QMC(opt) refers to calculations done
    with a fully optimized
    Slater determinant, which was VMC for the paramagnetic state (open circles), and DMC
    for the collinear state (green diamonds).
    The fully-optimized QMC calculation
    is done at $\Gamma$-point only, but at a 16-f.u. supercell. The
    DMC(PBE0) points are at 8 f.u., but are twist averaged over 8 twist
    values, therefore should have compareable finite-size errors.
    Accordingly, the ambient pressure DMC(PBE0) calculation agrees nearly within
    error bars with the fully optimized DMC(opt) calculation.
  }
\end{figure*}

% Pressure vs. volume.
\begin{figure}
  \includegraphics{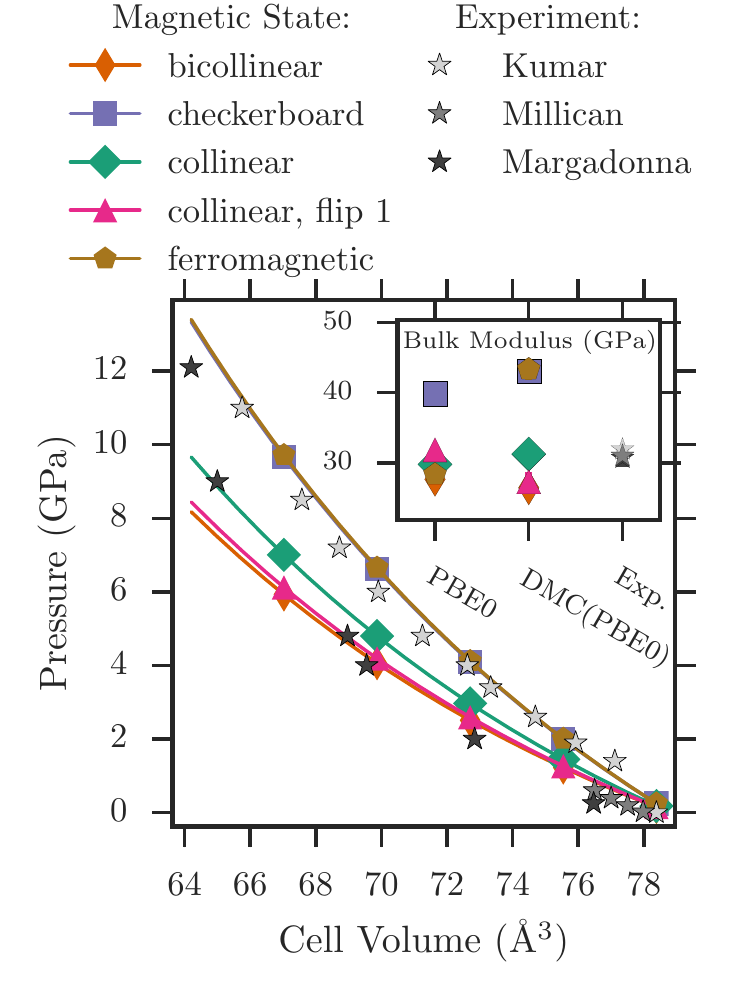}
  \caption{
    \label{fig:eos} 
    Pressure as a function of volume, computed through an equation-of-state
    fit to DMC(PBE0) data and from experiments by Margadonna \emph{et al}.
   ~\cite{margadonna_pressure_2009}, Kumar \emph{et al}.~\cite{Kumar}, and
    Millican \emph{et al}. ~\cite{millican_pressure-induced_2009}.  All
    points along the solid line come from the equation-of-state, and markers are
    added purely to distinguish the magnetic state. For the
    volumes considered here, regardless of spin ordering, $P(V)$ falls within
    the experimental spread.   Inset displays corresponding bulk modulus in
    units of GPa for PBE0, DMC(PBE0), and the experiments considered. The bulk
    modulus is strongly coupled to the magnetic state, and for the collinear
    state, DMC(PBE0) demonstrates excellent agreement with all the experiments
    considered. Bulk moduli computed from PBE are between 7 and 9 GPa,
    and are much more insensitive to the magnetic ordering (see
    Supplimental Materials for tabulated values). Lattice constants used were those of Kumar\cite{Kumar}.    
  }
\end{figure}

% Energies.

\section{Results}
  \subsection{Trial wave functions and ground state}

For the wave function in Eq.~\ref{eqn:sj}, there are many local minima both in preparing the Kohn-Sham orbitals using density functional theory and in optimizing the orbitals directly.
These minima correspond to different magnetic orderings of the Fe spins. The most relevant ones are presented in Fig.~\ref{fig:spinden}. 
We also included a type of paramagnetic state in which the up and down spin orbitals are constrained to be equal, but we found that state to be more than 0.5 eV/Fe higher in energy than any magnetically ordered wavefunction.
The ground state thus seems to require large local moments on the Fe atoms.

While it is known experimentally that FeSe does not have long-range ordering~\cite{ordering}, the calculations here enforce periodic boundary conditions on a relatively small cell and thus cannot describe long-range fluctuations of the magnetic order that might be the cause of loss of long-range order.
For the experimental crystal structure, the collinear magnetic ordering is the lowest in energy in our calculations and is observed to be the dominant short-range order experimentally~\cite{rahn_neutron_2015}.
The energetic cost of introducing a ``defect" into the magnetic order is quite small; we will discuss that aspect later.
Both the DMC(opt) and DMC(PBE0) approaches result in a rather large magnetic moment on the Fe atom.
For the collinear magnetic ordering we obtain a value of $\sim$ 3.4 $\mu_B$ for DMC(PBE0), and a slightly lower $\sim$ 3.1 $\mu_B$ for the fully optimized calculations. 
In both cases the magnetic moment is close to the atomic limit.

Between the two DMC approaches, the energy difference between different magnetic orderings is in agreement within stochastic errors, so there is good reason to believe that the cheaper DMC(PBE0) technique is accurate. 
In comparison to PBE calculations, which are the most common in the literature, the relative energies according DMC are quite different, including the lowest energy magnetic phase, which is the ``staggered dimer" configuration in DFT~\cite{Cao2015,Mazin2015,Tresca2015}, but turns out to be the collinear configuration in DMC.
It appears that hybrid DFT calculations in the PBE0 approximation obtain reasonably good magnetic energy differences in comparison to DMC; since this functional also produced the orbitals that gave the lowest FN-DMC energy, it may be capturing some of the correct physics for the magnetic properties of this material. 
However, the PBE0 functional predicts an insulating gap in agreement with previous work\cite{wu_modelling_2013} for FeSe for all magnetic orderings, 
in contrast to DMC and experiment.

\subsection{Crystal structure} \label{crystal_structure}

\begin{table*}
  \begin{ruledtabular}
    \begin{center}
      \caption{ FeSe optimal structural parameters with different computational methods. DFT calculations have been performed with the software package \textsc{QuantumESPRESSO}~\cite{qe} using a 10x10x10 k-points mesh, an energy cutoff of 75 Ry and norm conserving pseudopotentials for both Fe and Se.
      The variational Monte Carlo VMC(opt) results  are obtained at Gamma point only with the 16 f.u. FeSe supercell containing 32 atoms. 
      }
      \label{tab:geometry} 
      \begin{tabular}{ccccccc}
        \textsc{Source}
        & \textsc{Magnetic Ord.}
        & a 
        & c
        & $\overline{\textrm{FeFe}}$  
        & $z_\textrm{Se}$ \\   
      \hline
        DFT-PBE & paramagnetic & 3.6802 & 6.1663 & 2.6023 & 1.3862 \\
        DFT-PBE & collinear & 3.8007 & 6.2363 & 2.6966 & 1.4568 \\
        VMC & paramagnetic & 3.71(1) & 5.49(1) & 2.62(1) &   1.437(5) \\
        VMC & collinear & 3.72(1) & 5.68(1) & 2.63(1) & 1.56(1) \\
        experimental~\cite{Louca2010} - T 7 K &  & 3.7646(1) & 5.47920(9)  &  &  1.4622 \\
        experimental~\cite{Kumar} - T 8 K &  & 3.7685(1) & 5.5194(9)  & 2.6647(3) & 1.5879 \\
        experimental~\cite{margadonna_pressure_2009} - T 300 K &  & 3.7724(1) & 5.5217(1)  &  & 1.4759 
      \end{tabular}
    \end{center}
  \end{ruledtabular}
\end{table*}

Obtaining the correct crystal structure for FeSe is a major challenge, since the layers interact through non-bonded interactions. 
The $c$ lattice parameter in particular is affected by Van der Waals interactions and electron correlation plays a key role in determining the in-plane physics. 
The exotic behavior of FeSe superconducting properties under pressure gives another clue on the importance of structural variations in its description.  
A first-principles prediction of the lattice parameters is thus an important test of the description of this physics.
Since the DMC calculations are computationally costly, we limited our study to the tetragonal phase of FeSe.
Because the low temperature orthorhombic distortion is small\cite{margadonna_pressure_2009}, one might expect that its effect on the overall electronic structure is also small.
We leave such considerations to another paper.

The equilibrium lattice parameters of FeSe are presented in Table~\ref{tab:geometry}.
As mentioned in the previous sections, these results are obtained with a direct optimization of FeSe cell parameters with the VMC(opt) method. 
The in-plane FeSe properties should be well captured by QMC since the $a$ lattice parameter is in close agreement with experimental results (within $\sim$ 4 $\sigma$) independently of the chosen magnetic configuration.
Both collinear and paramagnetic wavefunctions shows also a general improvement with respect to DFT concerning the $c$ lattice parameter. 
This provides evidence of the accuracy in treating Van der Waals interactions with the QMC wavefunction, mainly achieved with the Jastrow factor\cite{vdw_qmc1,vdw_qmc2}.
The evaluation of the inter-plane $c$ distance might be affected by the dispersion along the z-axis, which we did not take into account in our supercell which always contains only one Fe plane. 
We check this dependency by performing a test structural relaxation on a FeSe supercell with 16 Fe atoms in two planes and 8 Fe atoms with only one plane considered in the supercell.
We find that the difference between the $c$ parameter obtained in the two configurations is negligible. 

The final internal parameter $z_\mathrm{Se}$ represents the height of the
selenium anion above the plane, and it has been experimentally demonstrated
\cite{Okabe2010} to be of key importance in determining superconducting
properties of iron-based superconductors in general.  
We collect all our calculations of $z_\mathrm{Se}$, as well as some experimental
results, in Fig. \ref{fig:zse}.
We find that both the magnetic state and the accuracy of the calculation 
have
an important effect on the prediction of this parameter.
At approximately the same level of finite size error, our two DMC calculations
agree very closely, determining that fixed node and basis set error is likely to
be unimportant. 
However, we found that $z_\mathrm{Se}$ is surprisingly sensitive to finite size
effects, both in the in-plane and out of plane directions.  
Given the supercells that we studied, we found a variation in $z_\mathrm{Se}$ of
approximately 0.05 \AA, depending on the twisted boundary conditions and
supercell.  
With experimental lattice parameters, our best estimate for $z_\mathrm{Se}$ is
thus 1.54(5)\AA, which is quite close to the experimental range.  
As we shall
see later, the properties of FeSe depend sensitively on $z_\mathrm{Se}$, so to
account for this uncertainty, we will consider properties as a function of
selenium height as well as pressure.

By fitting an equation of state previously used by Anton \textit{et al.}~\cite{anton_theoretical_1997} to our DMC(PBE0) energies as a function of volume, we extract the bulk modulus and the pressure dependence on volume $P(V)$, shown in Fig~\ref{fig:eos}.
The collection of ambient-pressure bulk-moduli results is reported in the inset of Fig.  \ref{fig:eos}, in units of GPa. 
For all these calculations, experimental lattice constants~\cite{Kumar} have been used. 
$P(V)$ and the bulk modulus show a strong dependence on the magnetic order.

While $P(V)$ has scatter between experiments, they are more consistent in the bulk modulus, so we base our comparisons of the theoretical calculations on the latter quantity. 
The DMC(PBE0) calculation demonstrates excellent agreement with all three experiments if the collinear magnetic ordering is imposed, but it is less close to experiment for the other magnetic orderings. 
Our PBE0 calculations are also in somewhat good agreement with DMC(PBE0), except a notable disagreement for the ferromagnetic ordering. 
On the other hand, PBE bulk moduli are significantly lower than both experiment and the other calculations, generally predicting bulk moduli between 7 and 10 GPa, depending only slightly on the magnetic ordering. 
Since the collinear ordering is also the lowest energy for DMC(PBE0), for the remainder of this article, we use the collinear equation of state to estimate the pressures that correspond to the volumes used in the calculations.

 \subsection{Interaction of structure and magnetism}
    \begin{figure*}
      \center
      \includegraphics{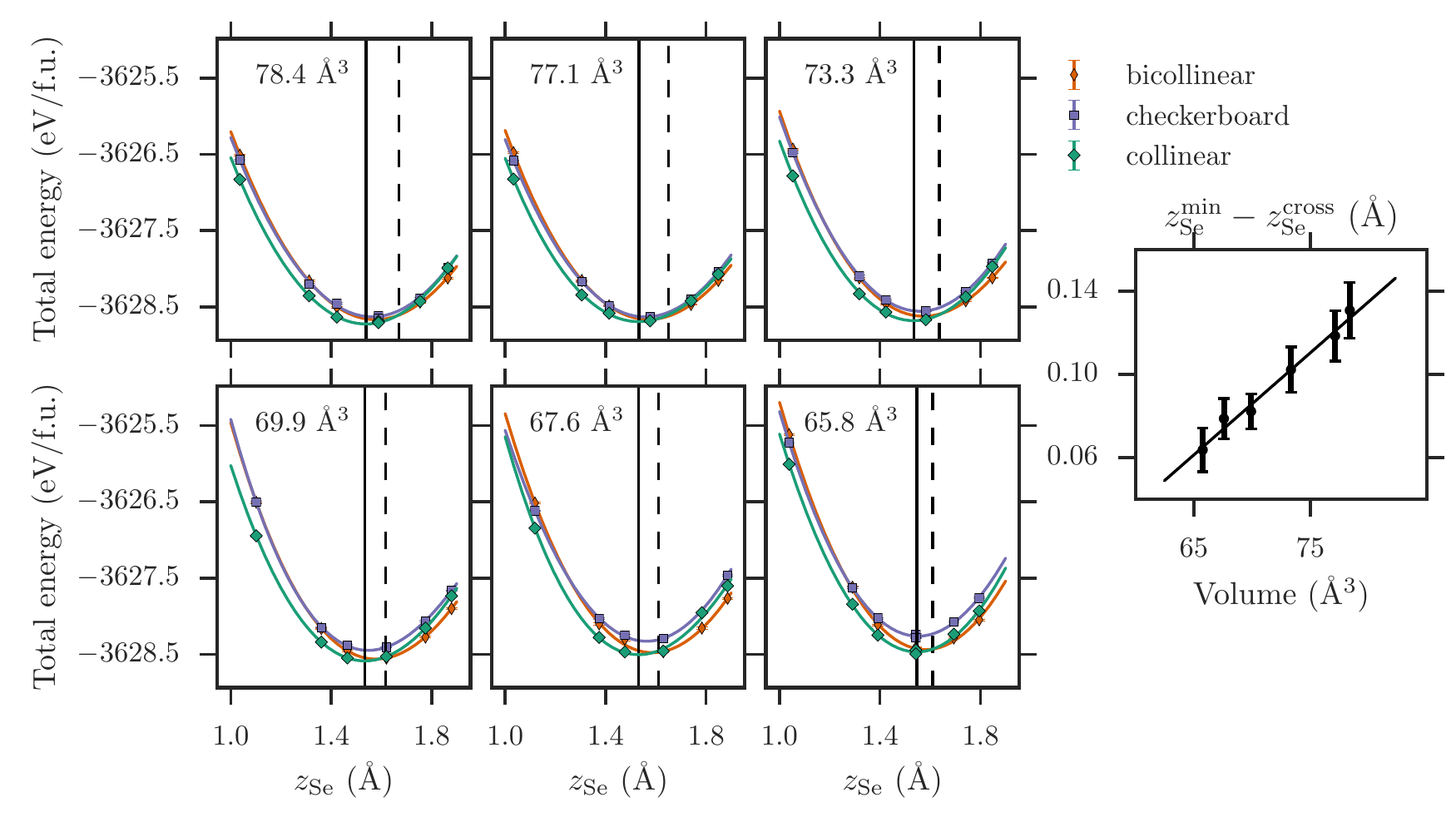}
      \caption{
        \label{fig:selandscape}
          DMC(PBE0) energies as a function of volume and selenium height $z_\mathrm{Se}$ for three of the magnetic orderings. 
          The solid vertical line represents the minimum of the collinear magnetic ordering, while the dashed vertical line represent the predicted crossing of the bicollinear and collinear energies. 
          These two points converge as the pressure is increased, as shown on
          the far right.
        }
    \end{figure*}

Fig~\ref{fig:selandscape} shows the interaction between pressure, magnetic ordering, and selenium height.
As has been found before~\cite{zSeSpinFluct2015}, the magnetic energies depend strongly on the selenium height, and this dependence changes with pressure.
For a given pressure, there are two points on the selenium height curves that are of substantial interest.
The first is the minimum energy (solid vertical line), which as can be seen in Fig~\ref{fig:zse}, does not change very much with pressure(or volume) in our calculations.
The second is the crossing point (dashed vertical line) between the collinear and bicollinear magnetic orderings, which are competing low-energy states.
This crossing point depends  on the pressure, and approaches the minimum energy point at higher pressures(lower volumes), as shown on the rightmost plot in the figure.

Another interesting feature visible in Fig~\ref{fig:selandscape} is that the checkerboard magnetic ordering intersects the bicollinear and collinear magnetic orders at zero pressure (78.4 \AA$^3$)  and large $z_\mathrm{Se}$, but there is a shift of the checkerboard curve to higher energies once pressure is applied.
The underlying physics of this effect will be discussed in Section~\ref{sec:magneto_orbital}. 

     \begin{figure*}
      \center
      \begin{tabular}{c}
        \includegraphics{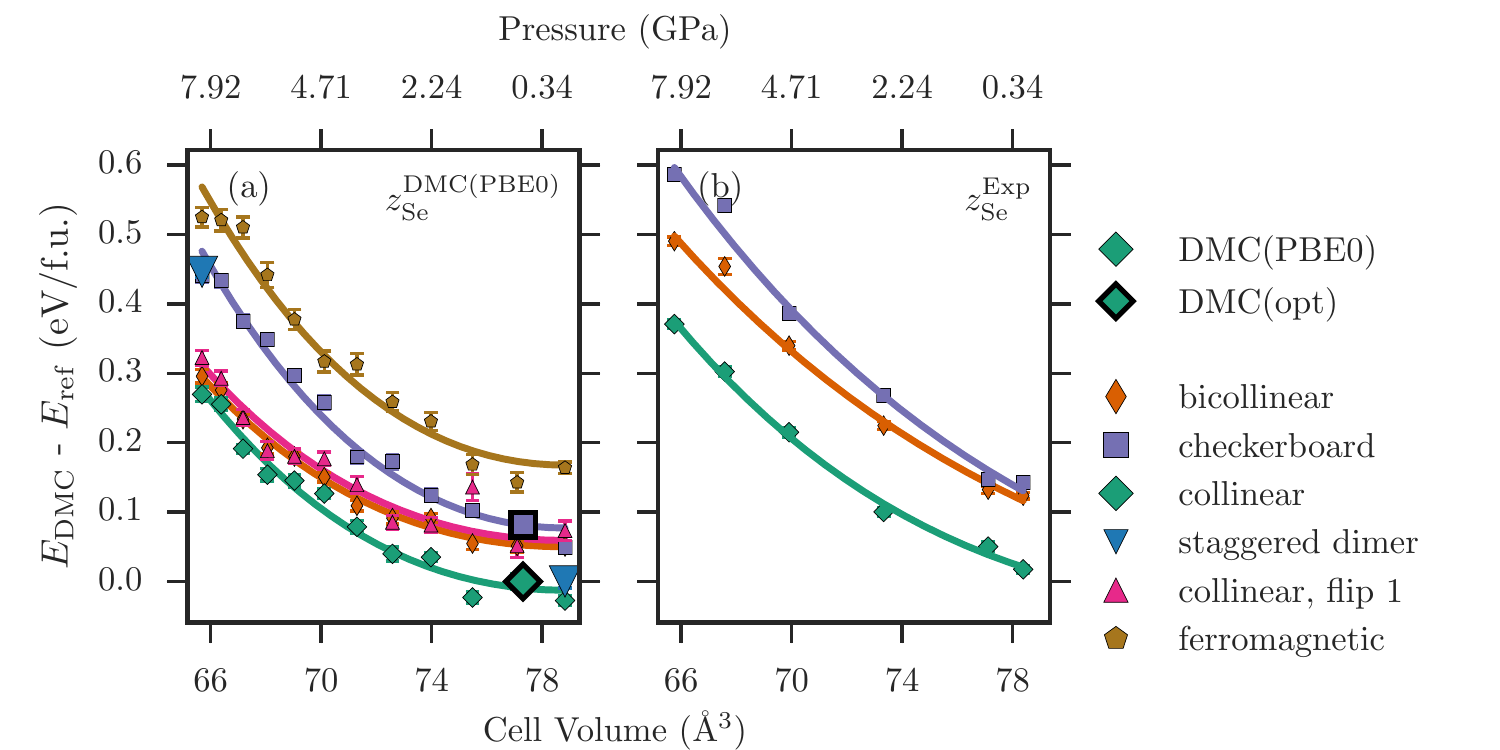}
        \\
        \begin{tabular}{cc}
           \includegraphics{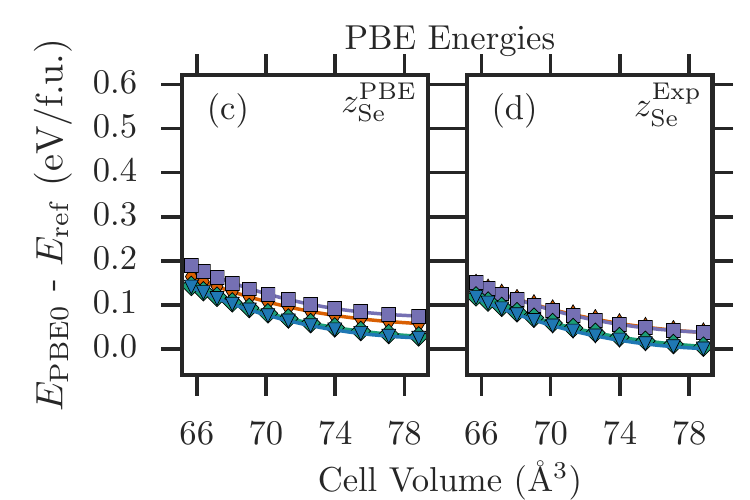}
           &
            \includegraphics{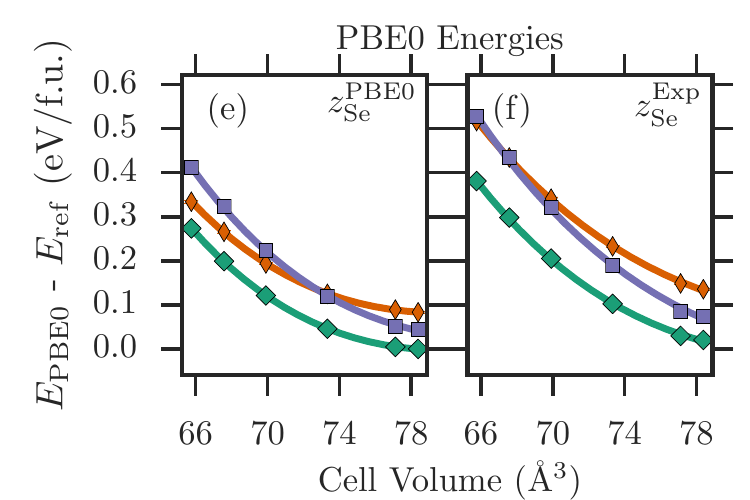}
           \end{tabular}
      \end{tabular}
      \caption{
        \label{fig:egydiff}
        For each calculation (QMC, PBE, and PBE0): (Right) Total energies for 8
        f.u. cell for various magnetic orderings, as a function of volume,
        choosing experimental~\cite{margadonna_pressure_2009} values of $z_\mathrm{Se}$. (Left) Same as right,
        but choosing optimized values of $z_\mathrm{Se}$. 
        For the top QMC plots, energies are referenced to the collinear energy at around 77 \AA$^3$. 
        The DFT calculations are referenced to the $z_\mathrm{Se}$ minimum energy for that type of calculation. 
        The DMC paramagnetic energies are $\sim$ 0.85 eV/f.u. higher than the reference collinear energy. 
      }
    \end{figure*}

Fig~\ref{fig:egydiff} shows a cut through the data in Fig~\ref{fig:selandscape} along the minimum energy $z_\mathrm{Se}$ (subfigure a), and the experimentally determined $z_\mathrm{Se}$ (subfigure b).
Along this cut we evaluated many magnetic orderings to establish a set of trends, and checked finite size errors by considering an 8 f.u. cell and 16 f.u. cell with twist averaging. 
Further information on finite size corrections are available in the Supplemental Material.
Under pressure, the checkerboard, ferromagnetic, and staggered dimer magnetic orderings rise in energy compared to the lowest energy collinear ordering. 
On the other hand, the stripe-like orderings, including the bicollinear and collinear orderings with defects converge with applied pressure.

From Fig~\ref{fig:egydiff} (bottom panels) is apparent the failure of PBE in capture this trend in FeSe energetics under pressure. 
Even with lattice constants fixed to experimental ones, the PBE energies of magnetically ordered states are quite different from the FN-DMC energies. 
In agreement with recent work, PBE does predict the staggered dimer as ground state.
Despite the failure of PBE0 in describing the conducting behavior of FeSe, the magnetic energies are reasonably close to the DMC results.

Given the data available to us, we can determine some properties that are robust to the finite size errors and uncertainty in $z_\mathrm{Se}$ in our calculations.
The first is that the relative energetics of magnetic orders changes strongly as a function $z_\mathrm{Se}$ and pressure.
In FN-DMC and PBE0, which would {\it a priori} be expected to be more accurate, the collinear and bicollinear orders become degenerate as a function of pressure for reasonable values of $z_\mathrm{Se}$.
According to FN-DMC, this effect is robust against $z_\mathrm{Se}$
variations, depending mainly on the change in the
relative magnetic energies as a function of pressure.

The energetic cost of reversing a single spin in the collinear ordered state, labeled ``collinear, flip 1'' in Fig~\ref{fig:egydiff}a, follows the bicollinear energy quite closely.
Because this cost decreases with pressure, we can surmise that magnetic fluctuations become more energetically available as pressure is increased.

    \subsection{Optical excitations and magnetism}
    
\begin{figure}
\includegraphics{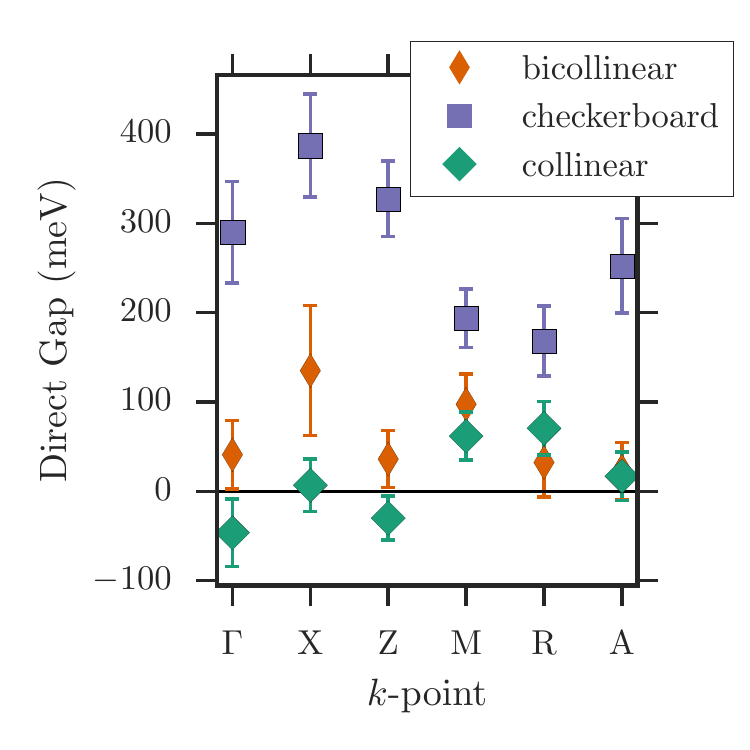}
\caption{
\label{fig:gap} 
Gap as a function of twisted boundary conditions calculated within DMC(PBE0).
For the stripe-like magnetic states, the gap is zero within statistical error.
The unit cell is a $2 \times 2$ supercell, expanded in the $x$-$y$ plane, shown
as one of the four outlined boxes in each of the spin densities of Fig.
\ref{fig:spinden}.
}
\end{figure}

The direct optical gap was calculated by promoting the highest energy orbital in the Slater determinant part of the trial wavefunction to the next excited state orbital. 
This constructs a wave function {\it ansatz} for an electron-hole excitation.
The results are shown in Fig~\ref{fig:gap}.
The resulting DMC(PBE0) energy relative to the DMC(PBE0) ground state is our estimation of the gap.
Interestingly, the DMC(PBE0) gap is within statistical uncertainties of 0 despite the fact PBE0 estimates a rather large gap, regardless of magnetic ordering.
Experimentally~\cite{shimojima_lifting_2014}, the gap is no more than 80 meV at any $k$-point, which is consistent with our results for the bicollinear and collinear magnetic ordering. 
Only the checkerboard state is gapped according to DMC(PBE0).

The charge degrees of freedom are therefore coupled to the spin degrees of freedom. 
According to these calculations, in FeSe there is a coupling between the mobility of charge and the spin ordering.
In the remainder of the paper, we will correlate these properties with 
those
of the ground state for different spin orderings.

 \subsection{Interaction of charge and orbitals with magnetism}
\label{sec:magneto_orbital}
    \begin{figure*}
      \center
      \begin{tabular}{c}
        \begin{tabular}{ccc}
          \includegraphics{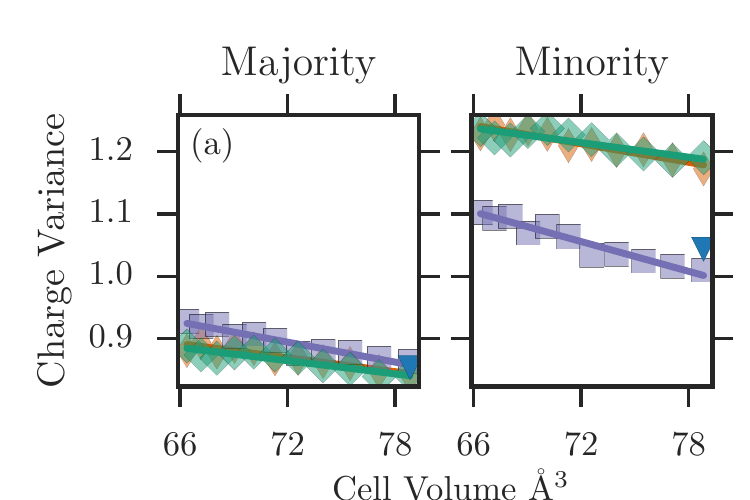}
          &
          \includegraphics{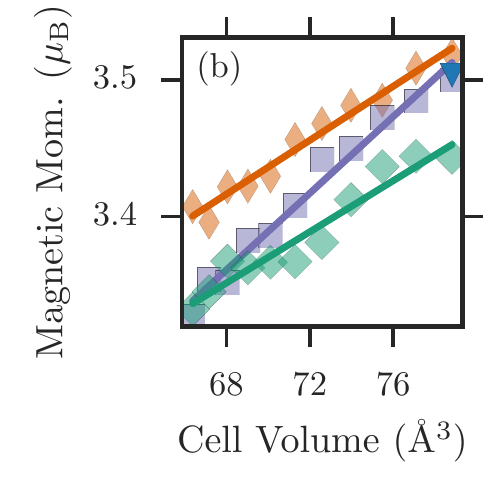}
          &
          \includegraphics{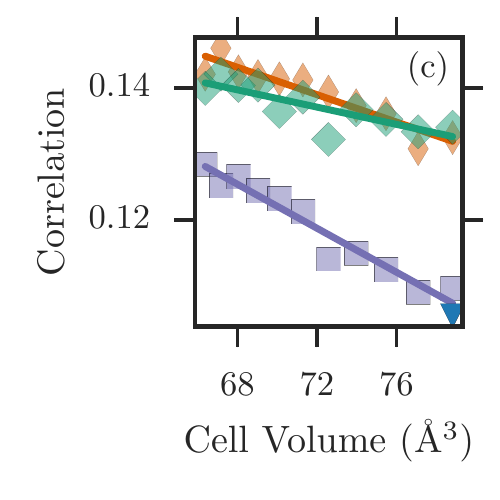}
        \end{tabular}
        \\
        \begin{tabular}{cc}
          \includegraphics{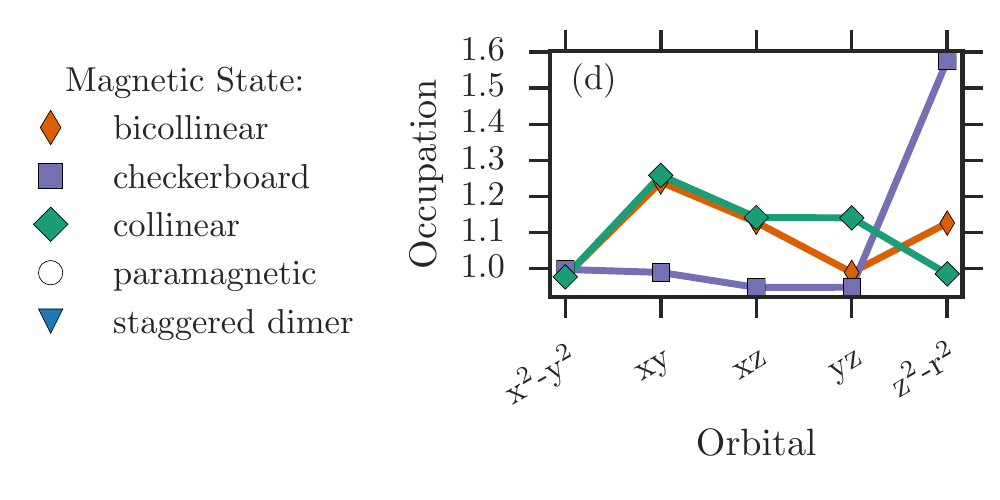}
          &
          \includegraphics{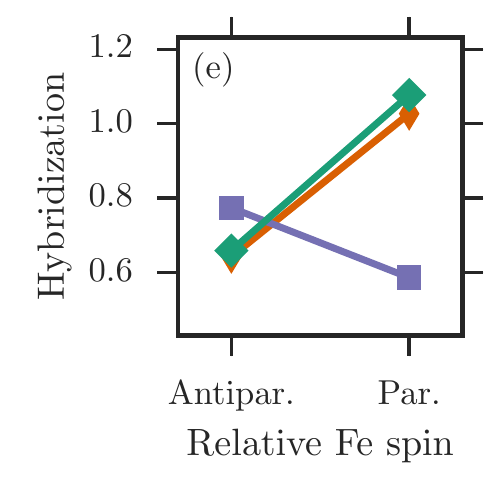}
        \end{tabular}
      \end{tabular}
      \caption{
        \label{fig:onebody}
  {\bf (a)} Charge variance for different magnetic orderings in the majority and minority spin channel, as a function of cell volume, illustrating that the minority spin channel is much more mobile, and additionally that the checkerboard ordering's electrons are more localized to a given iron site.  
  {\bf (b)} Magnetic moments inside each iron's Voronoi polyhedron.  
  {\bf (c)} Magnitude of on-site correlations, measured by
        $
          |\mathrm{Cov}(n^\uparrow,n^\downarrow)| / [\mathrm{Var}(n^\uparrow)\mathrm{Var}(n^\downarrow)]^{1/2}
        $.
 {\bf (d)} Single-particle orbital occupations of the iron $d$-states, measured by the on-diagonal terms of the 1-RDM. Note checkerboard's charge density points mostly out of the iron plane, while the other orderings point mostly within the plane. 
 {\bf (e)} Hybridization of different orderings, as measured by an average of the off-diagonal elements of the 1-RDM, broken down by interactions between antiparallel (Antipar.) aligned irons and parallel (Par.).
      }
    \end{figure*}
    
From the energetic properties, we note two classes of magnetic order in FeSe: ones which are stripe-like, and ones which are not stripe-like.  The stripe-like orderings converge in energy with pressure, while the checkerboard and staggered dimer pattern increases in energy relative to those orderings.
Similarly, the gap calculated in DMC(PBE0) distinguishes between different orderings, with metallic character in the stripe-like ordering.
In this section, we will make the following observations:
\begin{itemize}
\item Delocalized minority spin electrons are most affected by the spin ordering. 	
\item The one-particle orbitals are occupied differently depending on the spin ordering.
\item The above two effects change the effect of correlation in FeSe.
\end{itemize}
These effects combine to give a cartoon picture of the physics that explains the difference in pressure behavior between the magnetic orders.

\paragraph{Delocalization.} To address delocalization, we evaluate the charge compressibility of the Fe sites: $\langle (n_{i,\sigma}-\langle n_{i,\sigma} \rangle)^2 \rangle$, where $n_{i_\sigma}$ is the number of electrons within a Voronoi polehedron around the $i$th Fe site of spin $\sigma$. 
Larger values of the compressibility indicate more delocalized electrons. 
For a Fe atom with net $\uparrow$ spin, the $\uparrow$ electrons are labeled majority electrons and the $\downarrow$ minority, and vice versa for Fe atoms with net $\downarrow$ spin.
In Fig~\ref{fig:onebody}a, these results are presented.
For all magnetic orders, the majority spin is very similar and shows a low charge compressibility, while the minority spin is different between different magnetic orders, and its charge compressibility is larger.
From this, the minority electrons are more delocalized and their localization is affected by the magnetic order. 
For the stripe-like orders, the minority electrons are the most delocalized.

\paragraph{One particle orbitals.} In Fig~\ref{fig:onebody}d, we present the orbital occupation of the $d$ orbitals in different spin orderings. 
For stripe-like orderings, the $xy$, $xz$, and $yz$ orbitals are occupied, in
agreement with ARPES results~\cite{arpes}. 
On the other hand, the $3z^2-r^2$ orbital is occupied for the checkerboard ordering. 
This gives a simple explanation for the delocalization patterns:
The checkerboard pattern causes the minority spin to occupy the out of plane orbital, which would rise to an insulating state if it were the ground state.
This idea can be confirmed by checking the off-diagonal one-particle density matrix elements between Fe atoms with parallel and antiparallel net spins, in Fig~\ref{fig:onebody}e. The atomic orbitals are more hybridized between parallel spin Fe atoms for the stripe-like orders.
The charge degrees of freedom, which are mainly the minority spins from the Fe, interact strongly with the magnetic ordering.
This effect also interacts with the net magnetic moment and on-site correlations (Fig~\ref{fig:onebody}b and Fig~\ref{fig:onebody}c). 

 \subsection{A cartoon picture of FeSe}
\begin{figure}
      \includegraphics{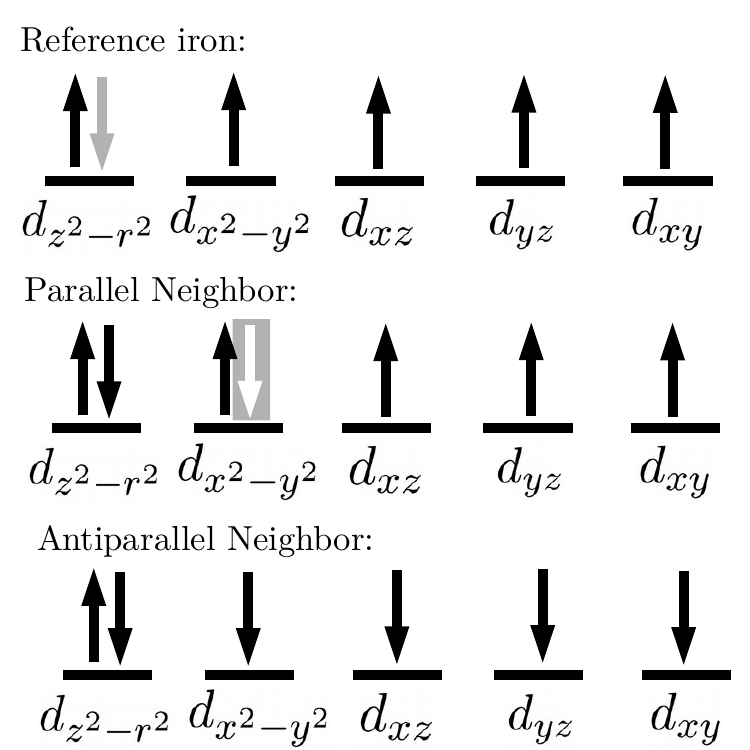}
      \caption{
        \label{fig:hunds}
        Diagrams depicting the occupations of the $d$-orbitals of a reference
        iron, one of its neighbors with parallel net magnetic moment, and
        another neighbor with antiparallel magnetic moment. The minority channel
        is spin down for the top two, and is spin up for the last. The minority
        electron on the reference atom is most likely to hop to a neighbor, for
        example, the greyed out down electron on $d_{z^2-r^2}$. It may easily
        hop to its parallel neighbor, which may fill its $d_{x^2-y^2}$ orbital
        as suggested by the grey box. It may not hop to any of the orbitals of
        the antiparallel neighbor, since the down spin channel is filled. Any
        hopping from the reference iron to its antiparallel neighbor must occur in
        the spin-up channel, which consequently violates Hund's rule for the
        reference iron.
      }
    \end{figure}

   A simple picture based on Hund's coupling can explain the energetics and other properties presented in the results section.  
    Hund's rules dictate that for an atom with a partially filled shell, we expect the electrons to have total spin $S$ that maximizes the multiplicity $2S+1$.  
    This is consistent with our computed magnetic moment, which find that the majority channel is mostly filled, bringing the moment to around 3.1-3.4 $\mu_B$. 
    The spin occupation of the $d$-states in a reference iron is diagrammatically shown in the top row of Fig. \ref{fig:hunds}. 
    Also due to Hund's coupling, the electron that is most likely to hop to nearby iron atoms would be the electron in the minority channel, to keep a large $S$. 
     As illustrated in Fig. \ref{fig:hunds}, this minority channel is already filled for neighboring irons that are antiparallel, so only majority spin electrons can hop to those atoms.
     Conversely, minority electrons can hop to neighboring parallel irons, since that spin channel is not filled. Thus, irons with parallel spins allow the minority electrons to delocalize, therefore decrease the kinetic energy. As seen in Fig. \ref{fig:onebody}d the magnetic ordering affects the occupation of the $d$ states, hence affects the labeling of the states in Fig. \ref{fig:hunds}, but the basic idea is unchanged.  

    While the minority spins require at least some parallel iron magnetic moments, the large localized magnetic moments also interact antiferromagnetically, leading to a competition between these two mechanisms. As a compromise, antiferromagnetic configurations with ferromagnetic chains emerge as the lowest energy configurations. 
    
    This picture unifies many of the observations from our calculations.  The
    checkerboard state is distinguished from the other states by its lack of
    parallel nearest neighbors, similar to how the ferromagnetic state is
    distinguished by its lack of antiparallel neighbors.  These two extremes are
    higher in energy, and are disfavored as pressure increases the importance of
    Fe-Fe interactions.  Because the checkerboard has no parallel nearest
    neighbor, its iron $d$-states are localized on the irons, leading to a low
    charge variance, and states that primarily occupy the $d_{z^2-r^2}$ orbital. 
    All stripe-like states have a combination of antiparallel and parallel
    nearest neighbors, and allow the electrons to delocalize along the irons
    chains, leading to higher correlations, higher variance, and more Fe-Fe
    hybridization. 
Although the staggered dimer ordering is energetically competitive at low pressures, its energy, charge variance, and magnetic moment are similar to checkerboard, and at high pressures, becomes energetically unfavorable just as the checkerboard ordering does. 
Although the staggered dimer does
    allow some delocalization between the dimered parallel spins, the itinerant
    spins are still trapped on the dimers, and therefore this state's energetics
    follow the checkerboard behavior at higher pressures.

    This competition of interactions sets up a fine balance between many qualitatively different magnetic configurations. 
    Parameters in the structure can tilt this balance one way or another, leading to a strong magneto-structural coupling. 
This is evident both from the strong magnetic dependence of the bulk modulus in
Fig. \ref{fig:eos}, as well as in Fig. \ref{fig:selandscape}, where
$z_\mathrm{Se}$ can exchange the ground state configuration between at least two
magnetic configurations. 
This logic can be straightforwardly  applied to iron telluride (FeTe), the non superconducting parent compound of FeSe.
This material has the ground state magnetic ordering at a $z_\mathrm{Te}$  around $1.75$ \AA, which implies that the $z_\mathrm{Te}$ should be decreased to force a crossover.
Therefore FeTe would superconduct if it were put in tensile stress, as it has been observed~\cite{han_superconductivity_2010}.

\section{Conclusion}

In summary, we have shown that QMC calculations can obtain an accurate description of the electronic structure of FeSe. 
The lattice constants, bulk modulus, and bandwidth are all very close to the experimental values and significantly improve over DFT calculations.
Our results are substantiated by the agreement between different and complementary QMC techniques employed. 
We showed that they yield sufficiently small statistical and systematic errors to study the relative energetics of different magnetic orders, which behave differently from those predicted by DFT.
The largest error in the calculations appear to be due to finite size supercells, which we checked to be small enough that the trends presented here are preserved.

As an outcome of the high-accuracy calculations, we have determined that collinear and bicollinear motifs become close in energy as pressure increases, while the checkerboard motif increases in energy with pressure. 
This behavior is correlated with delocalization of the minority electrons on the high-spin Fe atom. 
Collinear and bicollinear motifs allow for more delocalization, which increases in importance as the pressure is increased.
This delocalization effect is strong enough to change the occupation of atomic orbitals in FeSe depending on the magnetic ordering, so it is larger than the crystal field splitting of the orbitals. 
The spontaneous breaking of C$_4$ symmetry (or more properly S$_4$ symmetry)\cite{micheleFeSe,Maier2015}, is a result of this physics. 
Magnetic configurations which contain spin chains are thus favored over the whole considered range of pressure. 

From the above results, we can see that the magnetic degrees of freedom are strongly coupled with the charge and orbital degrees of freedom.
In a similar way, since the relative magnetic energies are dependent on the hopping of minority electrons from site to site, they are also strongly dependent on the structure.
There is thus both spin-charge and spin-structural coupling in this system. 
As one of us showed recently\cite{wagner_effect_2014,wagner_ground_2015}, the cuprates also show strong magneto-structural and magneto-charge coupling. 
One might speculate that both of these effects are necessary for high T$_c$ and it may prove fruitful to look for similar effects in proposed new superconductors.

\begin{acknowledgements}
We would like to thank the many people who have given useful suggestions and comments throughout this work.
They include 

	B.B. would like to thank the NSF Graduate Research Fellowship for funding. L.K.W. was supported by the U.S. Department of Energy, Office of Science, Office of Advanced Scientific Computing Research, Scientific Discovery through Advanced Computing (SciDAC) program under Award Number FG02-12ER46875. Computational resources were provided through the INCITE PhotoSuper and SuperMatSim programs.
S.S, M.C. and M.D. acknowledge computational resources provided through the HPCI System Research Project (Nos. hp120174 and hp140092) on K computer at RIKEN Advanced Institute for Computational Science, and on the HOKUSAI GreatWave computer under the project G15034.
M. C. thanks the GENCI french program to provide additional computer time through the Grant No. 2014096493.
\end{acknowledgements}

\bibliography{fese_combined}

\newpage

\section{Supplemental Material}

  \subsection{\label{sec:convergence} Convergence and validation}
   
    % Convergence plots.
    \begin{figure*}
      \begin{tabular}{c}
        \begin{tabular}{cc}
          \includegraphics{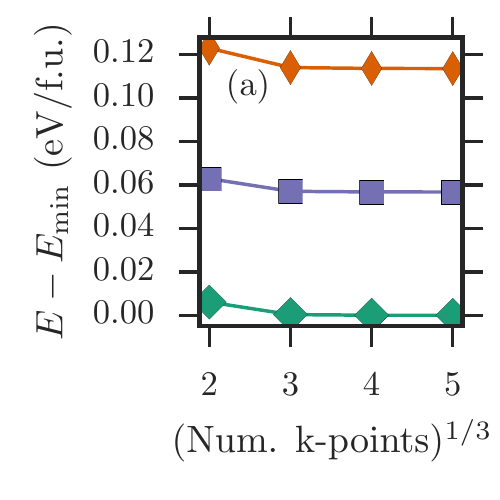}
          &
          \includegraphics{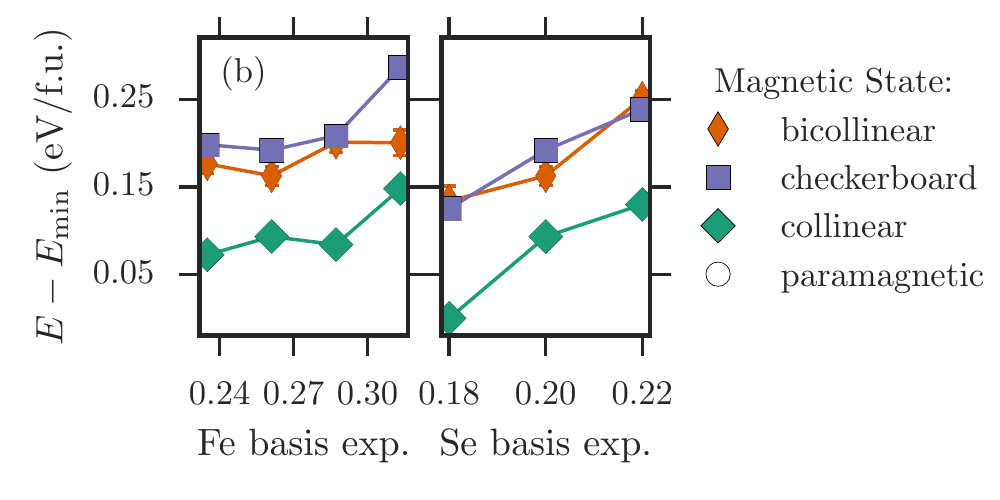}
        \end{tabular}
        \\
        \begin{tabular}{lll}
          \includegraphics{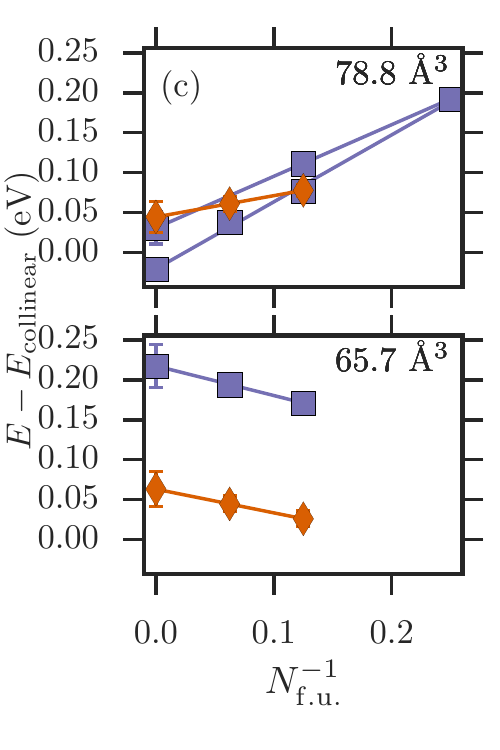}
          &
          \includegraphics{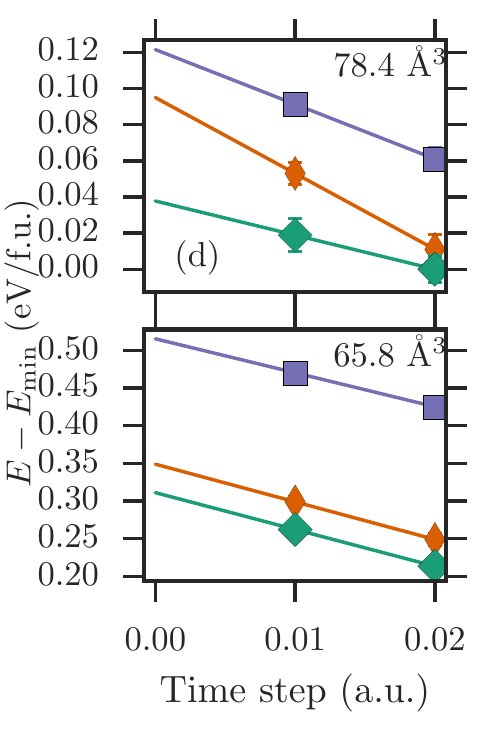}
          &
          \raisebox{\height}{
            \begin{tabular}{l}
              \includegraphics{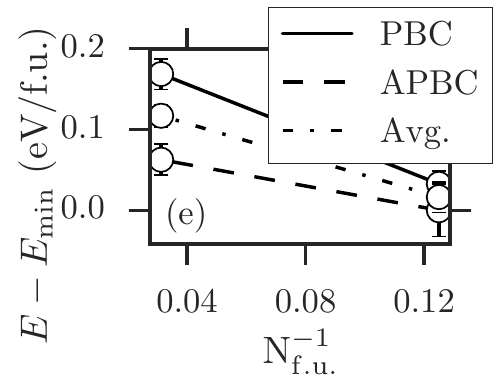}
              \\
              \includegraphics{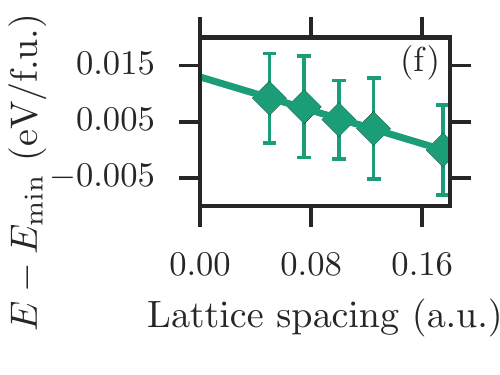}
          \end{tabular}}
        \end{tabular}
      \end{tabular}
      \caption{\label{fig:convergence} Convergence of DFT and
        DMC parameters. Axes labeled by $E - E_\mathrm{min}$ are energies
        relative to the minimum energy of the points in the subfigure. 
        (a) Convergence of k-point grid for DFT calculations for the 8 f.u.
        cell. A $4 \times 4 \times 4$ grid was used for all these calculations. 
        (b) Convergence of DMC(PBE0) basis set parameters for
        the 8 f.u. cell, at zero pressure. ``Fe basis exp" and ``Se basis exp"
        refer to the value of the smallest exponent in the basis. Values of
        around 0.26 and 0.2 were used for Fe and Se respectively. While in the
        latter case the total energy is decreasing, the energy differences
        are stable. 
        (c) Finite size extrapolation for DMC(PBE0) at the largest and smallest
        volumes considered, twist averaging over 8 boundary conditions. Energy
        is relative to the collinear energy for each size.  The two checkerboard
        lines reflect a finite size extrapolation in the $\hat z$ direction
        (upper line) and extrapolation in the $x$-$y$ plane (lower line). The
        full extrapolation will likely be near the center of the two line
        endpoints. 
        (d) Time step extrapolation for the DMC(PBE0) at 8 f.u. at the largest
        and smallest volumes considered. A value of 0.01 a.u. was adopted for all
        DMC(PBE0) calculations.
        (e) Finite size extrapolation for the DMC(opt) calculations, with
        periodic (PBC) and antiperiodic (APBC) boundary conditions, along with
        their average (Avg.). These benchmark calculations are performed in the paramagnetic
        phase and energies are corrected for one-body and two-body finite size errors.
        (f) Extrapolation of the lattice spacing used for the laplacian
        discretization for the LR-DMC, i.e.  DMC(opt), calculations. These benchmark
        calculations are performed in the collinear phase. A value of
        0.125 a.u. was adopted for all calculations.
      }
    \end{figure*}

    \begin{figure}
      \includegraphics{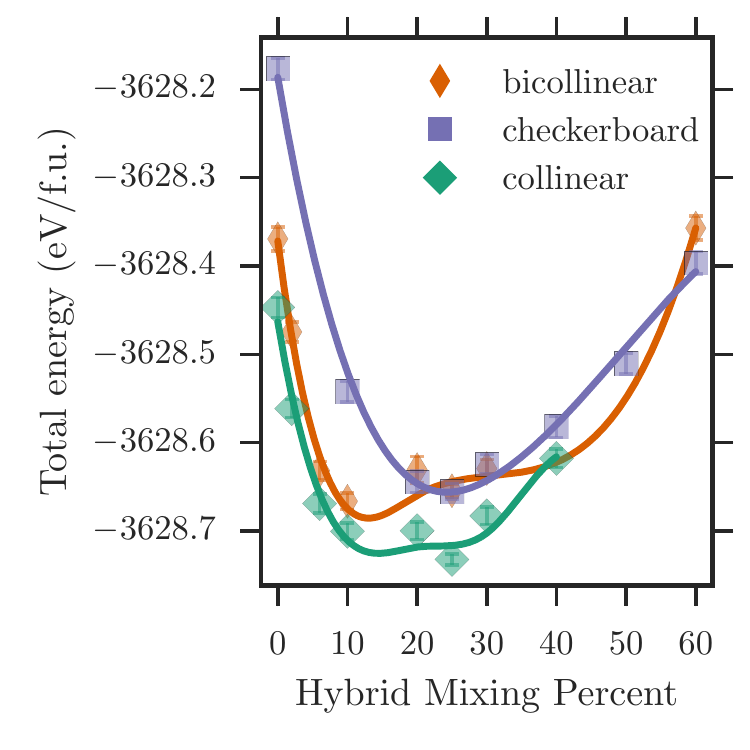}
      \caption{
        \label{fig:hybrid}
        DMC optimization of exchange correlation functional for the DFT
        calculations. The exchange-correlation function depends on $w$ as
        $
          E^w_\mathrm{xc}[n]
          =
          w E^\mathrm{HF}_\mathrm{x}[n]
          +
          (1-w) E^\mathrm{PBE}_\mathrm{x}[n]
          +
          E^\mathrm{PBE}_\mathrm{c}[n]
        $.
        Lines are a cubic spline interpolation as a guide to the eye.
      }
    \end{figure}
    
    In this section we present an extensive study of the convergence of the main
    parameters involved in our QMC calculations.  Within DMC(PBE0), the only
    parameter to be optimized in the Slater determinant is the amount of exact
    exchange $w$. The optimization of FN-DMC energy as a function of $w$ is
    presented in Fig~\ref{fig:hybrid}. As already mentioned in the main
    manuscript, we notice that the best FN-DMC is generally obtained with $\sim
    25\%$ of exact exchange for all magnetic configurations; this corresponds to
    the PBE0 density functional. The convergence of the FN-DMC energy with
    other parameters of QMC methods is reported in Fig~\ref{fig:convergence}.
    For the DMC(PBE0) method, in
    Fig.~\ref{fig:convergence}b, we present energy convergence as a function of
    Fe and Se basis set exponents by showing the behavior of the exponent of our
    most diffuse gaussian basis exponent as it gets more diffuse.  Benchmark
    calculations of DMC(PBE0) against the time step used in the projection are
    presented in Fig~\ref{fig:convergence}d.  We employed a time step of 0.01
    for all DMC(PBE0) calculations.  On the other hand, lattice regularized
    FN-DMC algorithm employed for DMC(opt) method (see next section) suffers
    from the lattice step error in Laplacian discretization. Convergence with
    lattice step is shown for the collinear configuration in
    Fig~\ref{fig:convergence}f. We used a lattice step of $0.125$ a.u. for all
    DMC(opt) calculations.

    We turn now our attention to finite size errors (FSE) which represent the
    main source of error in our QMC calculations.  
    We performed several DFT test
    calculations to determine the impact of one-body FSE. 
    For both 8 f.u. and 16
    f.u.  supercells (used for DMC(PBE0) and DMC(opt) methods respectively) a
    4x4x4 k-point grid is enough to obtain results converged within 1 meV
    independently from the density functional used, as shown in
    \ref{fig:convergence}a. 
    The same k-point grid in the PBE0 wavefunction is
    also sufficient to converge FN-DMC energies within the same threshold.  
    For the DMC(PBE0) calculations, we twist average over a set of 8 twist
    conditions~\cite{twist}, for unit cells ranging from 4 to 16 f.u., expanding
    the supercell in the $x$-$y$ plane, and $z$ direction (adding an additional
    layer). 
    The resulting finite-size extrapolations are depicted in
    Fig~\ref{fig:convergence}c. 
    The finite size extrapolation in the $z$
    direction is the checkerboard line above the other checkerboard line. 
    The true infinite size limit will likely lie in between these two
    extrapolations. 
    Although the finite size effects are relevant, they do not
    alter any conclusions of the main text. 
    Going from low to high pressure, the
    finite size errors amplify the change in energy differences between
    checkerboard and collinear. 
    Extrapolating with more than two points is prohibitively expensive
    for the bicollinear state, and so
    how finite size effects affects the pressure dependence of the energy differences between collinear
    and bicollinear is not clear, as the extrapolations are within error bars.
    However, it is certainly clear that bicollinear both remains very close
    in energy to the collinear state---even in the extrapolation---whereas
    checkerboard's energy certainly rises well above the other states.

    For DMC(opt), all calculations have been done with a 16 f.u. supercell.
    Structural optimization was performed with periodic boundary conditions only. All other calculations were instead 
    averaged between periodic (PBC) and fully antiperiodic (APBC) boundary conditions. Further finite size corrections 
    to DMC energies are obtained by adding one-body corrections estimated from fully converged DFT-LDA calculations 
    and two-body corrections evaluated within the KZK approach~\cite{kzk,silicon}. 
    In Figure~\ref{fig:convergence}e is reported the convergence of DMC(opt) energy in the paramagnetic phase as a function 
    of the system size after applying the corrections. 
    
    \subsection{QMC full orbital optimization}

    This section reports a detailed explanation of the QMC(opt) method.
    The wavefunction optimization procedure is carried out entirely with the variational Monte
    Carlo scheme. The initial Slater determinant is taken from a DFT calculation in the local
    density approximation (LDA) obtained with a built-in DFT package. The Slater determinant
    is expanded over a single-particle, localized gaussian basis set 
    $\chi_l^{\alpha}(\mathbf{r}-\mathbf{R}_{\alpha})$ 
    where $l$ is the basis set index and $\alpha$ is the nuclear index. 
    The number of gaussian functions employed for each angular 
    momentum channel is chosen to attain the best compromise between 
    basis set size and DFT(LDA) total energy. 
    The basis set we use is $(7s6p6d)/[3s2p2d]$ for Fe and $(5s4p4d)/[2s2p1d]$ for Se. 
    Following the notation, the round brackets contain the primitive basis. 
    The set of variational parameters involved in the orbital optimization can be highly reduced by
    building atomic basis contractions, i.e. linear combinations of the primitive orbitals. 
    The contracted basis employed for Fe and Se is contained in the square brackets. 
    The procedure for finding the optimal number of contracted orbitals will be explained in 
    detail in a manuscript is going to be published~\cite{hybrid_contr}.
    The Jastrow factor is multiplied to the DFT(LDA) determinant to build the final QMC ansatz. 
    For the orbital optimization scheme we choose a complex parameterization 
    of this term in order to include both charge and spin dependent terms.
    The argument of the exponential function in Eq.~\ref{eqn:sj} reads:
    \begin{equation}
      J(\mathbf{r}_1, \cdots \mathbf{r}_N) = \sum_{\alpha}^{N_{\textrm{atoms}}} \sum_j^{N} 
      g_{\alpha}^{1b}(\mathbf{R}_{\alpha} - \mathbf{r}) + \sum_{i \neq j}^N  g^{2b}(\mathbf{r_i},\mathbf{r_j})
      \label{eqn:jas}
    \end{equation}
    where $\mathrm{N}$ ($\mathbf{r}$) is the number (position) of electrons and $N_{\textrm{atoms}}$ 
    ($\mathbf{R}$) is the number (position) of atoms. Both terms on right-hand side are expanded 
    over the same localized gaussian basis set $\{\chi_l^{\alpha}(\mathbf{r}-\mathbf{R}_{\alpha})\}$.

    The first term on the right-hand side of Eq.~\ref{eqn:jas} is the electron-nucleus term:
    \begin{equation} 
      g_{\alpha}^{1b}(\mathbf{R}_{\alpha}-\mathbf{r}) = \sum_l
      G_{\alpha}^l  \chi_l^{\alpha}(\mathbf{r}-\mathbf{R}_{\alpha})
      \label{eqn:1body}
    \end{equation}

    The $g^{2b}$ term explicitly depends on two electron positions and it accounts 
    for both charge and spin fluctuations. It has the following form:

    \begin{align}
      g^{2b}(\mathbf{r},\mathbf{r}^{\prime})  &= \nonumber
      u(|\mathbf{r}-\mathbf{r}^{\prime}|) \; + \\ 
      &+ \sum_{{lm}}^{\alpha \beta} \nonumber
      C_{{lm}}^{\alpha \beta} \chi_l^{\alpha}(\mathbf{r}-\mathbf{R}_{\alpha}) 
      \chi_l^{\beta}(\mathbf{r}^{\prime}-\mathbf{R}_{\beta}) \; + \\ 
      &+ \sum_{{lm}}^{\alpha \beta} 
      S_{{lm}}^{\alpha \beta} \chi_l^{\alpha}(\mathbf{r}^{\uparrow}-\mathbf{R}_{\alpha}) 
      \chi_l^{\beta}(\mathbf{r}^{\prime\downarrow}-\mathbf{R}_{\beta})
      \label{eqn:2body}
    \end{align} 

    where the homogeneous term $u(\textrm{r}) = 0.5\textrm{r}/(1+\gamma\textrm{r})$ fulfills the 
    electron-electron cusp conditions for unlike-spin particles; the second term in the 
    right-hand side of Eq.~\ref{eqn:2body} 
    describes charge-charge interactions only, 
    while the last term explicitly includes 
    spin correlations and it is therefore suitable to describe 
    spin-spin fluctuations. We verified that the inclusion of spin-spin term yields a much ($\sim 0.1$ eV/Fe) 
    lower energy than considering the charge-charge term only. This Jastrow has therefore been used for all 
    QMC(opt) calculations presented in the main manuscript.

    The matrix elements $G_{l}^{\alpha}$,
    $C_{lm}^{\alpha \beta}$,
    $S_{lm}^{\alpha \beta}$ are taken as variational parameters
    and optimized during the energy minimization procedure. 

    The full wave function optimization requires several steps. At first the variational parameters
    of the Jastrow factor are optimized while the Slater determinant is kept fixed at DFT(LDA) level. 
    Afterwards the Slater determinant orbitals parameters (both the contracted orbitals coefficients 
    and the exponents of the gaussian basis functions) are included in the minimization procedure. 
    All the variational parameters of the Jastrow-Slater wavefunction are
    optimized by means of the stochastic reconfiguration method with
    Hessian accelerator, also known as ``linear
    method"~\cite{SR1,SR2,SR3}.

    Thanks to the adjoint algorithmic differentiation technique introduced in
    Ref.~\onlinecite{aad}, energy derivatives with respect to ionic 
    positions and lattice cell parameters can be performed in an efficient
    way (computational scaling $\sim N^3$ as for total energy evaluation), 
    by treating those quantities 
    on the same footing as the other wavefunction variational parameters. 
    This opens the way to obtain relaxed structures of periodic systems within
    a fully \emph{ab initio} variational Monte Carlo approach, as presented in the
    main manuscript.

    The energy gradients with respect to both wavefunction parameters and
    nuclear positions have always finite variance during the minimization procedure
    thanks to an efficient reweighting scheme~\cite{Attaccalite2008,Zen2013} which cures the
    divergence of variance in the proximity of the nodes.    

    In order to compute accurate 
    ground-state energies, once the trial wavefunction is fully optimized, we use a particular 
    implementation of the FN-DMC, the so-called Lattice Regularized diffusion Monte Carlo (LRDMC), introduced 
    in Ref~\onlinecite{lrdmc}. 
    This method provides a solution to the well-known issue of standard DMC when non-local pseudopotential are present as for 
    our FeSe calculations. The problem is solved by introducing an effective Hamiltonian 
    whose Laplacian (kinetic energy) term is evaluated on a discrete lattice. 
    Within LRDMC, the total energy is a variational upper bound 
    to the true ground state energy also in presence of a non-local pseudopotential. The trade-off is represented by 
    the Laplacian discretization error which is assessed in the previous section. 

    \subsection{Bulk moduli and magnetic moments}

    We report in this section the complete set of results concerning the bulk moduli (Table~\ref{tab:eos}) and the Fe magnetic 
    moments (Table~\ref{tab:mag_mom}) in FeSe. QMC calculations are compared with PBE and PBE0 outcome.

\begin{table}
  \caption{
    \label{tab:eos} Comparison of ambient-pressure bulk moduli from experiments
    and various calculations with various magnetic orderings. 
    All calculations were performed using the room temperature experimental
    structure.
  }
  \begin{tabular}{llc}
    \hline
    \hline
    source &           ordering &  Bulk Mod. (GPa) \\ 
    \hline
    Margadonna~\cite{margadonna_pressure_2009} & -- & 30.7(1.1) \\
    Millican~\cite{millican_pressure-induced_2009}   & -- & 31 \\
    Kumar~\cite{Kumar} & -- & 30.9(3) \\
    DMC(PBE0)  &  bicollinear    &      26.4(8)  \\
    DMC(PBE0)  &  checkerboard   &      43.1(8)  \\
    DMC(PBE0)  &  collinear      &      31.2(7)  \\
    DMC(PBE0)  &  collinear, flip 1 &   27(1)    \\
    DMC(PBE0)  &  ferromagnetic  &      43(1)    \\
    PBE0       &  bicollinear    &      27.6(1)  \\
    PBE0       &  checkerboard   &      40(1)    \\
    PBE0       &  collinear      &      29.7(2)  \\
    PBE0       &  collinear, flip 1 &   31.8(4)  \\
    PBE0       &  ferromagnetic  &      28.3(4)  \\
    PBE        &  bicollinear    &      7.03(3)  \\
    PBE        &  checkerboard   &      8.9(2)   \\
    PBE        &  collinear      &      7.3(1)   \\
    PBE        &  staggered dimer &     8.8(1)   \\
    \hline
    \hline
  \end{tabular}
\end{table}
    
\begin{table}
  \caption{
    \label{tab:mag_mom} Fe magnetic moments computed with DMC(PBE0) and
    DMC(opt) at different magnetic orderings. 
    For sake of comparison we also included PBE0 and PBE results. All calculations were performed using the room
    temperature experimental structure. }
  \setlength{\tabcolsep}{10pt}
  \begin{tabular}{lll}
    \hline
    \hline
    \textsc{Source} & \textsc{Magnetic Ord.}  & \textsc{Mag. Mom.}  ($\mu_B$) \\
    \hline
         DMC(PBE0) & bicollinear  & 3.518      \\
         DMC(PBE0) & checkerboard & 3.500      \\
         DMC(PBE0) & collinear    & 3.443      \\
         DMC(opt)  & checkerboard & 3.134(6)   \\
         DMC(opt)  & collinear    & 3.014(8)   \\
         PBE0      & bicollinear  & 3.492      \\
         PBE0      & checkerboard & 3.473      \\
         PBE0      & collinear    & 3.433      \\
         PBE       & bicollinear  & 2.653      \\
         PBE       & checkerboard & 2.332      \\
         PBE       & collinear    & 2.561      \\
         PBE       & stag. dimer  & 2.543      \\
    \hline
    \hline
  \end{tabular}
\end{table}

\end{document}